\documentclass[sigconf]{acmart}
\usepackage{threeparttable}
\usepackage{multirow}
\usepackage{enumitem}
\usepackage{subcaption}
\usepackage{mathtools}
\usepackage{listings}
\usepackage{bbm}
\usepackage{makecell}
\usepackage{subcaption}
\usepackage{amsmath}
\usepackage{diagbox}
\usepackage{hyperref}
\usepackage{multicol}
\usepackage{listings}
\usepackage{xcolor}
\usepackage{xspace}
\usepackage{ifthen}
\usepackage{cleveref}
\usepackage{microtype}

\newcommand{\projName}{\texttt{polymer}\xspace}
\newcommand{\ProjName}{\texttt{Polymer}\xspace}
\newcommand{\Volvo}{Volvo\xspace}
\newcommand{\spapi}{\texttt{spapi}\xspace}

\newboolean{showcomments}
\setboolean{showcomments}{true} 
\ifthenelse{\boolean{showcomments}}{
  \newcommand{\nbc}[3]{
    {\colorbox{#3}{\bfseries\sffamily\scriptsize\textcolor{white}{#1}}}%
    {\textcolor{#3}{\sf\small$\blacktriangleright$\textit{#2}$\blacktriangleleft$}}}
}{
  \newcommand{\nbc}[3]{}
}

\definecolor{mygreen}{rgb}{0.15, 0.65, 0.25}
\definecolor{myred}{rgb}{0.72, 0.13, 0.13}

\newcommand{\aka}{\hbox{\emph{aka}}\xspace}

\definecolor{codegreen}{rgb}{0,0.6,0}
\definecolor{codegray}{rgb}{0.5,0.5,0.5}
\definecolor{codepurple}{rgb}{0.58,0,0.82}
\definecolor{backcolour}{rgb}{0.95,0.95,0.92}

\lstdefinestyle{mystyle}{
    commentstyle=\color{codegreen},
    keywordstyle=\color{magenta},
    numberstyle=\tiny\color{codegray},
    stringstyle=\color{codepurple},
    basicstyle=\ttfamily\footnotesize,
    breakatwhitespace=false,         
    breaklines=true,                 
    captionpos=b,                    
    keepspaces=true,                 
    numbers=left,                    
    numbersep=5pt,                  
    showspaces=false,                
    showstringspaces=false,
    showtabs=false,                  
    tabsize=2,
    morekeywords={assert}
}
\lstdefinelanguage{my-yaml}{
  keywords={spec, containers, name}, 
  keywordstyle=\color{blue}\bfseries,
  moredelim=[is][commentstyle]{||}{££}, 
  identifierstyle=\color{black},
  sensitive=false,
  comment=[l]{\#},
  commentstyle=\color{olive}\ttfamily,
  stringstyle=\color{orange}\ttfamily,
  morestring=[b]',
  morestring=[b]"
}

\usepackage[linesnumbered,ruled]{algorithm2e}
\SetAlgorithmName{Procedure}{Procedure}{List of Procedures}
\AtBeginDocument{%
  \providecommand\BibTeX{{%
    \normalfont B\kern-0.5em{\scshape i\kern-0.25em b}\kern-0.8em\TeX}}}

\makeatletter
\def\@ACM@copyright@check@cc{}
\makeatother
\setcopyright{cc}
\setcctype{by}
\copyrightyear{2025}
\acmYear{2025}
\acmConference[FSE Companion '25]{33rd ACM International Conference on the
Foundations of Software Engineering}{June 23--28, 2025}{Trondheim, Norway}
\acmBooktitle{33rd ACM International Conference on the Foundations of
Software Engineering (FSE Companion '25), June 23--28, 2025, Trondheim,
Norway}\acmDOI{10.1145/3696630.3728494}
\acmISBN{979-8-4007-1276-0/2025/06}

\begin{document}

\title{Polymer: Development Workflows as Software}


\author{Dhasarathy Parthasarathy}
\affiliation{%
  \institution{Volvo Group}
  \country{}
}
\author{Yinan Yu}
\affiliation{%
  \institution{Chalmers University of Technology}
  \country{}
}
\author{Earl T. Barr}
\affiliation{%
  \institution{University College London}
  \country{}
}


\begin{abstract}
Software development builds digital tools to automate processes, yet its
initial phases, up to deployment, remain largely manual. There are two reasons:
Development tasks are often under-specified and transitions between tasks
usually require a translator.  These reasons are mutually reinforcing: it makes
little sense to specify tasks when you cannot connect them and writing a
translator requires a specification. LLMs change this cost equation:  they can
handle under-specified systems and they excel at translation. Thus, they can
act as skeleton keys that unlock the automation of tasks and transitions that
were previously too expensive to interlink. We introduce a recipe for writing
development workflows as software (\projName) to further automate the initial
phases of development. We show how adopting \projName at \Volvo, a large automotive
manufacturer, to automate testing saved 2--3 FTEs at the cost of two months to
develop and deploy. We close with open challenges when \projName{izing} 
development workflows. \looseness=-1


\end{abstract}



\maketitle
\section{Introduction}
\label{sect:intro}

Software Engineering (SE) in large organizations is a complex
undertaking~\cite{fuggetta2014software}. All SE organizations aim to operate
this complex process at high cadence while delivering quality software
products. Achieving this goal takes tremendous effort because it requires
tackling SE process complexity as a whole. For instance, in a large automotive
company like \Volvo, developing a portfolio of 3-5 vehicle platforms involves
hundreds of engineers developing and integrating thousands of SW components
with thousands of mechatronic and physical components. The combinatorial
explosion of interactions means that management and engineering teams can lose
track of each other's roles and responsibilities. The resulting \emph{awareness
gap} increases waste, slows development, and reduces quality.\looseness=-1


\vspace{1mm}
\textbf{Automation manages complexity better}: Consider a typical SE process with phases such as specification, implementation, verification/validation, deployment, and operation. Interestingly, the tail-end phases already employ a powerful approach for handling complexity --- treating entire phases like deployment or operations as software problems~\cite{47176} and automating workflows by writing them as software. Thanks to the widespread adoption of DevOps, aided by a deep tech stack, engineers code complex workflows that continuously build, test, deploy, and monitor SW products, some at global scale catering to millions of users. How can SW-defined DevOps workflows handle high complexity? We offer a few explanations: \looseness=-1

\begin{itemize}

    \item SW-defined workflows are more formal, making them clearer, bridging the awareness gap between stakeholders.
     
    \item Parts of SW-defined workflows can be executed, obviating manual work and providing faster feedback with less coordination. 
    
    \item Workflow instrumentation and logging permits close observation and measurement, which enables organizations to control complex SW processes with leaner management.
    
    \item SW-defined workflows can be simulated, enabling continuous refinement to be more reliable, observable, and faster.

\end{itemize}

Given these benefits and the success of SW-defined workflows in handling DevOps complexity, we ask: ``Can we left-shift the idea of SW-defined workflows and replicate this success in earlier SE phases?''. The \projName vision centers on answering this question. \looseness=-1

\begin{figure*}[ht!]
    \centering
    \includegraphics[width=1.0\textwidth]{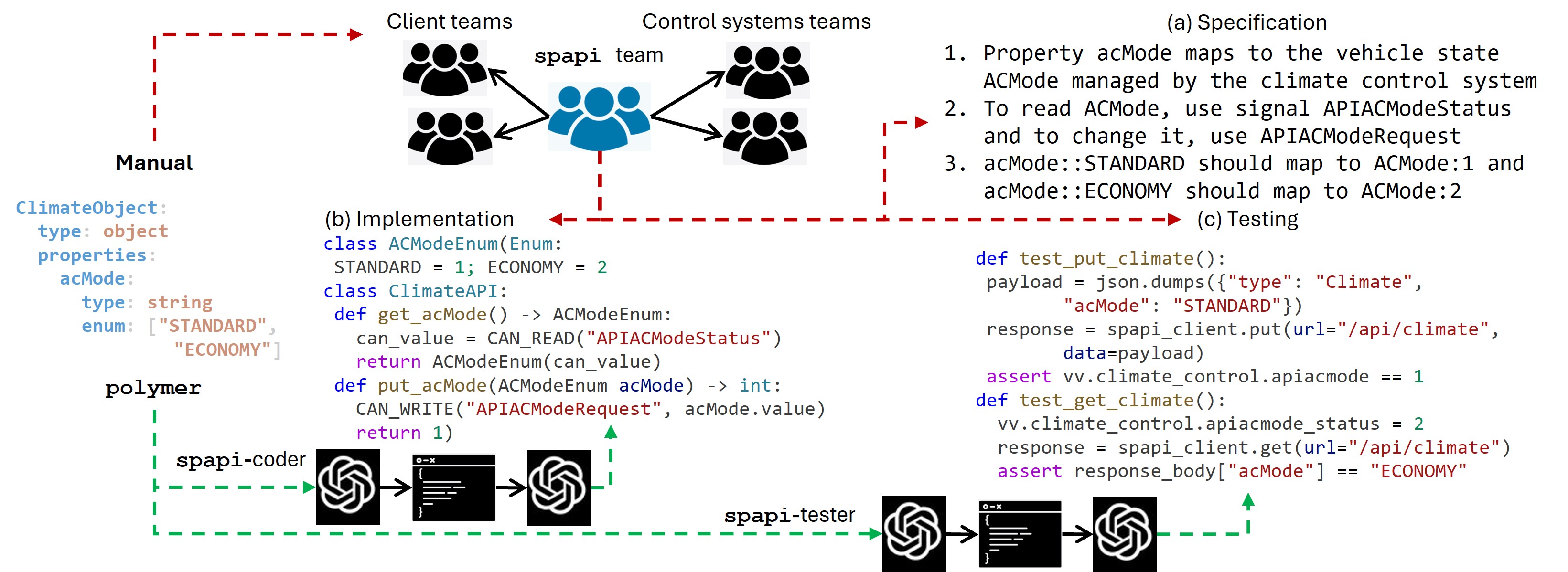}
    \caption{\spapi is representative of a complex SE process involving continuous coordination during specification, implementation, and testing. Applying \projName, this workflow becomes SW-defined and automatic, reducing development complexity and saves thousands of person hours. Note: examples have been sanitized to protect proprietary information.\looseness=-1}
    \label{fig:spapi-endpoint-example}
    \vspace{-10pt}
\end{figure*}

\vspace{1mm}
\textbf{The \projName Challenge}: We model an SE workflow as a graph where nodes represent tasks and edges represent transitions, or information flows, between tasks. Within this structure, the level of specification varies. Product tasks, which produce tangible outcomes like software or documentation, tend to be better specified. Control tasks such as coordination or hand-off tend to be under-specified. Transitions are usually as well specified as the tasks they interconnect. As workflow complexity and awareness gaps grow, so does the level of under-specification. Applying \projName involves automating as many tasks and transitions as possible, so it requires addressing under-specification directly. For tasks, automation entails formalizing inputs and outputs, and making it executable, which is expensive. Likewise, automating transitions requires translators capable of parsing, interpreting, and processing the information flows, which is equally expensive. These costs can trigger a vicious \emph{under-specification cycle} where high costs and awareness gaps lead to vague specifications. This hinders automation and, with feature growth, increases complexity. The increased workload leaves no room for improving specifications, perpetuating the cycle. Often, extra human intervention is required to keep this cycle from spiraling.\looseness=-1

\vspace{1mm}
How did DevOps break out of this cycle and achieve mature SW-defined workflows? We argue that this success is primarily due to the standardization of tail-end SE tasks. Diverse applications, from media streaming to vehicle embedded systems, have implicitly joined forces to standardize tasks such as CI/CD, regression testing, infrastructure provisioning, and observability. This standardization creates high demand and reuse potential, justifying years of collective effort to progressively automate tail-end tasks and transitions.\looseness=-1

\vspace{1mm}
However, if we shift left from the tail-end into the earlier SE phases, we risk
slipping into the long tail of workflow diversity, where
 some workflows are so domain-specific as to be
unique. For such cases, the cost of conventional automation remains prohibitive.
Why, then, do we advocate adopting \projName in early phases of development? We
do so because of the advent of a crucial tipping factor --- Large Language
Models (LLMs). An LLM's general task solving ability dramatically changes the
economics of handling under-specification and automating task chains,
with less concern about standardization, formalization, or
executability. \looseness=-1 

\vspace{1mm}
\textbf{LLMs as  ``skeleton keys''} unlock automation opportunities: 
\begin{enumerate}
    \item LLMs are task-agnostic~\cite{brown2020languagemodelsfewshotlearners, bubeck2023sparksartificialgeneralintelligence, schlangen2024llmsfunctionapproximatorsterminology}, and can be adapted to solve diverse tasks using lean prompting. This is often cheaper than implementing every task as code. \looseness=-1
    
    \item LLMs handle semi-structured and semi-formal inputs, reducing, and sometimes eliminating, data wrangling~\cite{narayan2022can, jaimovitch2023can}. \looseness=-1
    
    \item LLM outputs can be structured and adhere to syntactic constraints~\cite{DBLP:conf/icml/Beurer-Kellner024, ugare2024syncodellmgenerationgrammar, park2024grammaraligneddecoding}, allowing them to be composed with conventional, deterministic functions in pipelines. \looseness=-1
    
    \item LLM have a flexible front-end, supporting natural language, programmatic~\cite{DBLP:journals/corr/abs-2310-03714, DBLP:journals/pacmpl/Beurer-Kellner023}, and multi-modal interactions, enabling diverse SE stakeholders to solve tasks with them.
    
    \item LLMs can serve as a flexible back-end, addressing SW tasks, data tasks, and even some HW tasks~\cite{DBLP:journals/corr/abs-2405-02326, Makatura2024Large}.
\end{enumerate}

Based on these observations, we claim that \textit{LLMs are skeleton keys for development workflow automation}. That is, LLMs enable automation with far less effort than conventional methods, making it feasible to automate, \aka \emph{polymerize}, task sequences that were previously prohibitively expensive to automate. Building upon this claim, we propose a simple recipe for applying \projName and reducing the cost of handling SE complexity: Take a development workflow and (a) identify tasks and information flows suitable for automation, (b) write these tasks and flows as \emph{code}, invoking LLMs where needed and using conventional automation otherwise. Taking concrete steps towards \projName, we contribute the following: \looseness=-1

\begin{enumerate}
    \item We present a real-world case of test workflow as software. Developed in two months, \spapi-tester is a \projName process that automates testing \spapi --- a vehicle web server produced by \Volvo --- and saved the effort equivalent to 2-3 FTEs. \looseness=-1

    \item We discuss a real-world case of implementation workflow as software. Being piloted at \Volvo, \spapi-coder aims to \projName-ize \spapi implementation. \looseness=-1
    
    \item We lay out a research agenda for \projName enabled by LLMs as skeleton keys for automation.
    
\end{enumerate}

\section{A motivating example}


For our discussion, we use the development workflow of \spapi (\Cref{fig:spapi-endpoint-example}) as a running example. It represents our target application domain --- a complex workflow marked by under-specification, awareness gaps, and continuous coordination. A production web server deployed in \Volvo trucks, \spapi offers RESTful APIs allowing mobile applications to access and modify vehicle state. For example, the \texttt{/speed} endpoint is used to read the vehicle's speed, while \texttt{/climate} adjusts the cabin climate settings. \looseness=-1

\vspace{1mm}
\textbf{Specification is a coordination vortex}. While specification is always a challenging aspect of development, workflows with multiple stakeholders can demand overwhelming coordination. With \spapi, it often takes several discussions before requirements engineers even recognize the need for an endpoint like \texttt{/climate} with a property \texttt{acMode} (see \Cref{fig:spapi-endpoint-example}). Using their domain knowledge, they link this endpoint to the climate control system. However, the organizational and disciplinary separation between \spapi and climate engineers creates awareness gaps, driving up the cost of maintaining robust specifications. Incomplete or outdated vehicle state documentation forces manual coordination to match \texttt{ACState} with \texttt{acMode}. Moreover, because vehicle states are accessed via Controller Area Network (CAN) signals, \spapi architects must also identify or define the associated read and write signals (\texttt{APIACModeStatus} and \texttt{APIACModeRqst}), requiring further coordination with the signal definition team. After extensive coordination, the mapping between a property, signal, and state is recorded using natural language text (\Cref{fig:spapi-endpoint-example}(a)). Since \spapi comprises hundreds of APIs and thousands of attributes, \Volvo spends thousands of hours on specification. Even so, clarity and stability in specification is hard to achieve. \looseness=-1

\vspace{1mm}
\textbf{Implementation can fall into the coordination vortex}. Given clear specifications, translating them into a working implementation can be straightforward in processes like \spapi. For an endpoint like \texttt{/climate}, developers align data types between the property \texttt{acMode}, and state \texttt{ACState}, before using CAN signals \texttt{APIACMode*} to read and write vehicle state (\Cref{fig:spapi-endpoint-example}(b)). However, due to constant changes in property, signal, and state definitions, specifications inevitably degrade, forcing developers to coordinate with stakeholders, adding hundreds of hours of overhead. \looseness=-1

\vspace{1mm}
\textbf{Testing is also claimed by the vortex}. Ambiguity and frequent specification changes inevitably affect testing. For \spapi, testers manually write and maintain test cases to ensure that endpoints comply with the specified interface and correctly map API properties to vehicle states. For example, they write a test case for a \texttt{PUT} method that sends \texttt{acMode:STANDARD} and checks if \texttt{ACState} changes to \texttt{1} (\Cref{fig:spapi-endpoint-example}(c)). However, since they rely upon the same set of degrading specifications, testers are forced to spend hundreds of additional hours coordinating with requirements engineers, developers, and close stakeholders to resolve differences. \looseness=-1 

\vspace{1mm}
Effective coordination and precise specifications are essential in SE, yet real-world workflows like \spapi continually struggle with deficiencies in both. The usual response is to introduce more management, restructure organizations, and enforce process discipline --- measures that are expensive and offer no guarantees. Conventional automation might help, but the unique nature of these workflows and the need for extensive change make it costly and unlikely to be prioritized. How can we break out of this stalemate and improve the process? \Volvo is piloting \projName using LLMs as skeleton keys. \looseness=-1

\section{Test workflow as code}

A typical test workflow involves translating specifications into test cases and executing them to assess quality. In real-world settings, as seen in the \spapi case, it needs to address the complexity arising out of under-specification and awareness gaps between multiple stakeholders. Left shifting by applying \projName, \Volvo addresses these issues by developing \spapi-tester\footnote{{\label{ftn:repo}} Note: These are sanitized illustrations to protect proprietary information.} (\Cref{lst:spapi-tester}), a SW-defined workflow that automates the test process. Testing \spapi starts with reading specifications (e.g., mapping \texttt{acMode:STANDARD} to \texttt{ACState:1}), followed by determining mocks and test inputs and integrating test cases into a regression suite --- a process that takes 2-3 FTEs. Rather than building a costly semantic parser~\cite{kamath2018survey}, \spapi-tester uses an LLM as a translator via the \texttt{dspy} framework~\cite{DBLP:journals/corr/abs-2310-03714}. Invoking LLM calls through declarative Python modules using \texttt{dspy.signature}, it efficiently implements \texttt{PropertyToStateMapper} to parse specifications and output corresponding mappings as a \texttt{dict}. \looseness=-1
\lstset{basicstyle=\ttfamily\footnotesize}
\begin{lstlisting}[language=Python,float,breaklines=true, belowskip=-2.5 \baselineskip, caption={\spapi-tester: \projName for \spapi testing.},captionpos=b,label={lst:spapi-tester}]
class SpapiTester:
    def __init__(self):
        self.prop_to_state = dspy.TypedChainOfThought(PropertyToStateMapper)
        self.gen_test_obj = dspy.TypedChainOfThought(TestObjGenerator)
        self.method_tester = MethodTester()
    def forward(self,api_obj,mapping_spec,method_type):
        # LLM as parser
        mapping = self.prop_to_state(api_obj, mapping_spec)
        # LLM as fuzzer
        test_obj = self.gen_test_obj(mapping)
        # Conventional automation
        test_case = self.method_tester(test_obj)
        return test_case
\end{lstlisting}

\vspace{1mm}
Upon translation, the task of writing a test case remains. The conventional option is to use a REST API fuzzer~\cite{golmohammadi2023testing} but, to avoid the cost of selecting and adapting one, \spapi-tester uses the skeleton-key LLM itself as the fuzzer. It defines \texttt{TestObjGenerator} to (a) take the LLM-assembled mapping between endpoint properties like \texttt{acMode} and corresponding vehicle states like \texttt{ACState}, and (b) asks the LLM to choose a corresponding pair like \texttt{acMode:STANDARD} and \texttt{ACState:1} as test objects. Finally, there remains the step of using the generated objects to write the test case and test the endpoints and these steps are quite simple to conventionally automate. 

\vspace{1mm}
Applying \projName, \spapi-tester polymerized one information flow (property to state mapping) and one product task (writing test cases) within a short span of 2 months, automating 2-3 FTEs worth of effort. Following a successful pilot to confirm its quality and utility \cite{wang2025automating}, \Volvo is currently deploying \spapi-tester. Though this brings valuable gains, the coordination burden remains largely intact.\looseness=-1 

\section{Implementation workflow as code}
\label{sec:spapi:imp}

Encouraged by the relative ease with which \projName automates testing, we consider left shifting further by applying it to the implementation workflow. Naturally, the complexity of translating a specification into a working implementation is of a higher order. Take \spapi, where implementing an endpoint like \texttt{/climate} with a property like \texttt{acMode}, requires synthesis of methods \texttt{get\_acMode} and \texttt{put\_acMode}. The workflow involves three key steps --- discovering the corresponding vehicle state \texttt{ACState}, discovering corresponding CAN signals \texttt{APIACMode*} to read/write state, and using this information to implement get and put methods. Discovering correspondences in a state of under-specification, as we previously noted, requires intense coordination. \Volvo estimates that completing these three steps for hundreds of \spapi APIs requires 15-20 FTEs. Applying \projName, and leveraging the skeleton key LLM, \spapi-coder\footnotemark[1] (\Cref{lst:spapi-coder}) bypasses manual coordination and automates endpoint implementation by treating it as a case of program synthesis. \looseness=-1

\lstset{basicstyle=\ttfamily\footnotesize}
\begin{lstlisting}[language=Python,float,breaklines=true, belowskip=-2.5 \baselineskip, caption={\spapi-coder: \projName for \spapi implementation.},captionpos=b,label={lst:spapi-coder}]
class SpapiCoder:
    def __init__(self):
        self.signal_synthesizer = SignalSynthesizer()
        self.signal_retriever = SignalRetriever()
        self.prop_synthesizer = dspy.TypedPredictor(PropertySynthesizer)
        self.template_examples = fetch_examples()    
    # LLM + conventional program synthesis
    def synthesize_signals(self, state_def):
        rd_signal, wr_signal = self.signal_synthesizer(state_def)
        self.signal_retriever.insert_signal(rd_signal)
        self.signal_retriever.insert_signal(wr_signal)
    # RAG-based program synthesis
    def synthesize_property(self, property, k=5):
        top_k_signals = self.signal_retriever.k_relevant(property["description"], k)
        return self.prop_synthesizer(top_k_signals, self.template_examples)
\end{lstlisting}

\vspace{1mm}
Using skeleton-key LLMs, \spapi-coder addresses all three dimensions ~\cite{gulwani2017program} of program synthesis --- (a) user intent, (b) a search space, and (c) a search technique. To synthesize REST APIs like \texttt{/climate}, \spapi-coder uses OpenAPI specifications to capture user intent. The \texttt{synthesize\_property} function takes the OpenAPI property definition as input to synthesize its \texttt{GET} and/or \texttt{PUT} methods. Since endpoint methods access state by reading and writing CAN signals, the search space includes all possible CAN signals corresponding to vehicle states. In control applications, it is often possible to extract the definition of a state like \texttt{ACState} from the code. The \texttt{synthesize\_signal} function takes the definition of a state like \texttt{ACState} to synthesize read/write \texttt{APIACMode*} signals. To search over this space and synthesize endpoint methods, \spapi-coder uses Retrieval-augmented Generation (RAG). By loading all signal definitions into a vector database, and using the property definition as the query, \texttt{synthesize\_property} searches for the $k$ most relevant signals and feeds it into an LLM call defined by \texttt{PropertySynthesizer}. Using the most relevant signals, and some few-shot examples, this LLM call synthesizes get and put methods for accessing state. This bridges all disciplinary gaps, fixes under-specification, and polymerizes endpoint synthesis, eliminating the coordination vortex. \looseness=-1

\vspace{1mm}
Leveraging \projName and the skeleton-key LLM, \spapi-coder polymerizes a workflow consisting of three tasks and the vortex of information flows that interconnect them. Unlike \spapi-tester, which is currently being deployed, \spapi-coder is currently in a pilot phase. \Volvo is seeing encouraging signs that it helps automate \spapi implementation, potentially streamlining 15-20 FTEs worth of effort.\looseness=-1

\section{Open challenges}

Despite its adoption at \Volvo, saving 2-3 FTEs in one case and promising to save 15-20 FTEs in another, challenges to unlocking \projName's potential remain. Here are a few. \looseness=-1

\vspace{1mm}
\textbf{What to \projName?}: \spapi turns out to be a successful showcase of \projName because (a) there is a clear case of inflated complexity, and (b) there is a pre-existing decomposition which can be SW-defined and polymerized using skeleton-key LLMs. Such clarity may be absent in other real-world workflows, so organizations need to carefully assess whether \projName is suitable for a given workflow and which of its tasks and transitions should involve LLMs. How does one efficiently and effectively make this decision? What risk versus gain calculus does one use to ease this decision-making?

\vspace{1mm}
\textbf{Back to the future}: The adoption of \projName reintroduces an old problem: companies must validate the \projName processes they implement. How can one ensure that \projName processes reliably produce SW artifacts at scale, and build sufficient trust among stakeholders? What do test cases for \projName processes even look like? How does the test oracle problem change if the system under test is a development workflow? How do we develop workflow automation oracles, and how much human involvement do they require? \looseness=-1


\vspace{1mm}
\textbf{Agile change management}: On the one hand, \projName will revolutionize change management. Since \projName processes are executable and can be simulated, they allow management to experiment with org structures at low cost. It will help them answer questions like --- How do we ensure that the organization is resilient enough to withstand market shocks while also being responsive to new opportunities? On the other hand, adopting \projName will itself require substantial change management. The technology is still new to many stakeholders, and it is still maturing. How can organizations progressively transition to \projName while adequately addressing these limitations, and also continuing to deliver customer value?

\vspace{1mm}
\textbf{Bridging the engineer-MBA gap}: Today, in cases like \spapi, engineers focus on development while managers concentrate on market demands, cost efficiency, and strategic planning. Given the intertwined nature of management and development, can \projName help break this strict separation to better integrate management and development activities in a workflow? In such a converged state, can it ensure that decision-making seamlessly incorporates both technical and business insights to identify and meet market needs? \looseness=-1

\vspace{1mm}
We call upon the SE community to engage in research and collaborative efforts to address these open questions.

\vspace{-5mm}

\section{Related Work}

Development workflows as software offer a novel approach to Software Process Improvement (SPI)~\cite{kuhrmann2016software}. Unlike traditional process-focused SPI methods, \projName treats processes as software systems, making monitoring, feedback, iterative refinement easier at scale. \projName is a new form of Model-Driven Engineering (MDE) \cite{da2015model} that leverages LLMs, as skeleton keys, to seamlessly formalize and model existing stakeholder processes rather than requiring the disruptive adoption of a new process and the domain-specific language that usually accompanies it. Further, as shown in our example and in recent studies~\cite{grand2023lilo, vella2024synergistic, liventsev2023fully}, LLMs are well-suited for producing software. \ProjName offers a systematic framework for integrating LLMs with conventional automation to achieve this. An important framework that is parallel to our proposal is multi-agent systems~\cite{he2024llm}, which orchestrates collaborative LLM agents to solve SE tasks. While autonomous solution discovery and evaluation are powerful, \projName provides a complementary path for cases where task decomposition and solution spaces are better known. Not only can this speed up task solving, it can also ease the evaluation of solutions. Additionally, the deliberate design of development workflows, merging LLM use with conventional automation, applies the core principles compound AI systems \cite{compound-ai-blog} to automate SW development.\looseness=-1

\vspace{-5mm}

\section{Conclusion}

We have presented \projName: a promising new approach for automating development workflows. By decomposing workflows and combining LLMs with conventional automation, \projName writes development tasks and transitions as code.  Applying \projName at \Volvo achieved practical cost savings. A wider application to fully realize its potential raises open questions that require further research. We invite the community to join us in answering them. \looseness=-1


\bibliographystyle{ACM-Reference-Format}
\bibliography{bibliography}



\end{document}